\documentstyle{article}
\begin{document}

\begin{center}
{\large\bf Kawasaki dynamics and equilibrium distributions in simulations of 
phase separating systems}
\end{center}

\begin{center}
{Claudio S. Shida and Vera B. Henriques \\
Instituto de F\'{\i}sica, Universidade de S\~{a}o Paulo, \\
Caixa Postal 66318, 05389-970 S\~{a}o Paulo-SP, Brazil}
\end{center}

\begin{abstract}
We compare equilibrium probability distributions obtained from Monte Carlo
simulations for different spin exchange dynamics with the exact Boltzmann
distribution for the fixed magnetization Ising model on small lattices. We
present simple arguments and numerical evidence in order to show that
"efficient" Kawasaki exchange does not lead to Boltzmann equilibrium
distributions and that nearest-neighbour equal spin-exchange must be
considered. Arbitrary distance opposite spin interchange is indicated as an
alternative to obtaining the full phase diagram with phase separation.

\end{abstract}

Equilibrium properties under phase separation are a problem of current
interest, both in simple\cite{1} and complex solutions, such as polymer
mixtures\cite{2}, microemulsions\cite{3} or different types of solutions of
amphiphile molecules in water which present liquid crystal phases or local
structure such as micelles, vesicles and membranes\cite{4}. Coexistence in
simple solutions is analogous to the problem of liquid - gas coexistence and
obtaining full concentration-temperature phase diagrams is still a subject
of discussion in the literature\cite{5}. As to complex solutions, even very
simple lattice model systems which include finite multi-site molecules\cite
{2}\cite{6} have no analytical solution beyond Flory mean- field\cite{7} and
simulations have been one of the main tools of investigation\cite{8}. In
solutions of amphiphilic molecules or copolymers one is not only interested
in describing the thermodynamic phase diagram, but one also wishes to obtain
the morphology of the different phases or microphases. In fact, one is often
not sure whether different morphologies represent different phases. For this
reason, simulations are most often carried out at constant concentration,
that is, in the canonical, rather than in the grand- canonical ensemble\cite
{9}. To obtain the equilibrium structures one must deal with the long time
relaxation problem of nucleation and growth, an object of innumerous studies
in the case of simple liquids\cite{10}. We therefore found it adequate to
explore the limitations of constant-density simulations\cite{11} in relation
to equilibrium properties in the critical and phase separating region. In
order to be able to compare simulation and analytical results, although our
interest lies in complex solutions, we chose to study the case of the Ising
model at constant magnetization (or, equivalently, the lattice gas model at
constant density), which corresponds to fixed relative concentrations of a
model mixture of molecules of the same size. This model has been studied
often enough in order to obtain growth behaviour\cite{10} and, more
recently, critical dynamic exponents\cite{12}, but we found very little
reference to recent studies of its equilibrium properties\cite{15}. In order
to be able to compare with analytical results, we focus attention on the
two-dimensional case and equal relative concentrations, the equivalent of
the magnetic model at zero field . The well-known Kawasaki spin exchange
model\cite{14}, originally formulated as a kinetic model associated to a
master equation, is the appropriate tool for the case. To our surprise, we
found out that a straight-forward application of the spin-exchange algorithm
in the form suggested by the main text-books on simulations\cite{15} results
in non-Boltzmann equilibrium distributions. The necessary search for
efficiency in simulations of the spin exchange dynamics, has lead, among
others\cite{16}, to the idea of selecting opposite pairs of spins in order
to save computer time\cite{17}. However, this time-saving procedure in the
choice of trial exchange moves does not fulfill the detailed balance
condition and therefore produces incorrect equilibrium distributions\cite{18}
, as we show below.

Kawasaki dynamics in simulations is presented in textbooks\cite{14} as an
algorithm in which one must find a pair of opposite sign nearest neighbours
and try to exchange them according to some detailed balance criterion, for
example, the Metropolis prescription\cite{19}. Exchanging spins of the same
sign does not alter the configuration and would apparently mean a waste of
time. The difficulty in obtaining convergence of the thermodynamic
properties of the model in the large $L$ limit to the known analytic forms
lead us to compare equilibrium properties of the model obtained through
different procedures for the trial moves. As long as one is not interested
in dynamic properties\cite{10}, the choice of kinetic model should be
irrelevant. We have applied different density conserving algorithms to the
zero magnetization Ising model on a square lattice. Three of them are
presented in this paper: ($i$) interchange of nearest neighbours
selected randomly among pairs of unlike spins, ($ii$) exchange of
nearest neighbours selected randomly, independently of sign (this includes
moves which do not alter the configuration of spins) and ($iii$)
interchange of spins selected randomly among pairs of unlike spins, but
independently of distance. These three procedures lead to the distribution
curves for the spin configurations shown in fig 1, where the exact Boltzmann
distribution is also shown, since it is easily calculated for such small
lattice ($L=4$). The striking fact is that method ($i$) leads to a
non-Boltzmann equilibrium distribution. The reason for this result is, as a
matter of fact, quite simple, as we try to illustrate below.

Consider the configuration (a) of fig 2 for an $L=5$ lattice. According to
the nearest neighbour opposite spin exchange algorithm, the probability that
it turns into configuration (b) during the simulation is $\frac 16e^{-\beta
\Delta E}$, because there are six pairs of opposite sign, whereas the
probability that configuration (b) turns back into (a) is $\frac 18$, since
there are now eight pairs of opposite sign. This does obviously not satisfy
detailed balance. It is easy to think of other examples for which a bias
towards higher energy configurations is present.

On the other hand, the same figs show that detailed balance will be obeyed
if one exchanges nearest neighbours independently of sign, or random
distance pairs of opposite sign. The probabilities are, respectively, $\frac
1{50}e^{-\beta \Delta E}$ and $\frac 1{46}e^{-\beta \Delta E}$ for process $
a\rightarrow b$ and $\frac 1{50}$ and $\frac 1{46}$ for $b\rightarrow a$.
This explanation for our result is apparently obvious, but we found no
reference whatsoever to this problem in the literature\cite{20}.

One might hope that the distortion would be smaller for larger lattices. To
show that this is not the case, we compare energy distributions obtained
from the different algorithms for $L=10$ and $L=4$ in fig 3. The Boltzmann
probability distribution has not been calculated in the first case, but the
comparison of the three algorithms is clear. In the case of the opposite
pair exchange, the curve is slightly distorted at the edges for $L=4$,
whereas for $L=10$ the whole curve is shifted towards higher energy values.

The effect of the distortion of the equilibrium distribution on
thermodynamic quantities is quite drastic. To exemplify, we show this in
relation to energy and specific heat in figs 4 and 5. Since procedures 
($i$) and ($ii$) involve long relaxation and correlation times,
much longer runs are needed in these cases, in relation to case ($iii$),
 as indicated in the figure captions. Fig 6 shows that proper care was
taken in order to assure that these effects were taken into account. It is
quite clear from figs 4 and 5, that while convergence to the expected
behaviour of an infinite lattice is achieved in the case of algorithms 
($ii$) and ($iii$), the first algorithm would indicate a much
smaller transition temperature, as a consequence of the fact that those
transition probabilities make ''higher temperature configurations'' more
probable.

Perhaps one should recover the original formulation of Metropolis and
cowor- kers\cite{19}, in which the transition probability is written as a
product of two factors: the a priori probability which links two states a
and b, which depends on the choice of movement one uses for the updates, and
the exponential factor $e^{\beta (E_b-E_a)}$. For simulations in liquids,
this is the usual prescription\cite{21}. For spin systems, however,
different updating procedures are of current use, with no check on the full
transition probability.

We have presented results for the critical case, for clearness of argument.
However, the short relaxation and correlation times presented by the random
distance exchange point to the possibility of obtaining the full phase
coexistence diagram from the specific heat peak \cite{22}, using finite size
analysis, with small numerical cost. It thus stands as an alternative to
other methods proposed in the literature\cite{11}.

Acknowledgments. We acknowledge support by Funda\c {c}\~{a}o de Amparo \`{a}
Pesquisa do Estado de S\~{a}o Paulo (FAPESP).

\newpage
\begin{center}
{\large\bf Figure Captions}
\end{center}

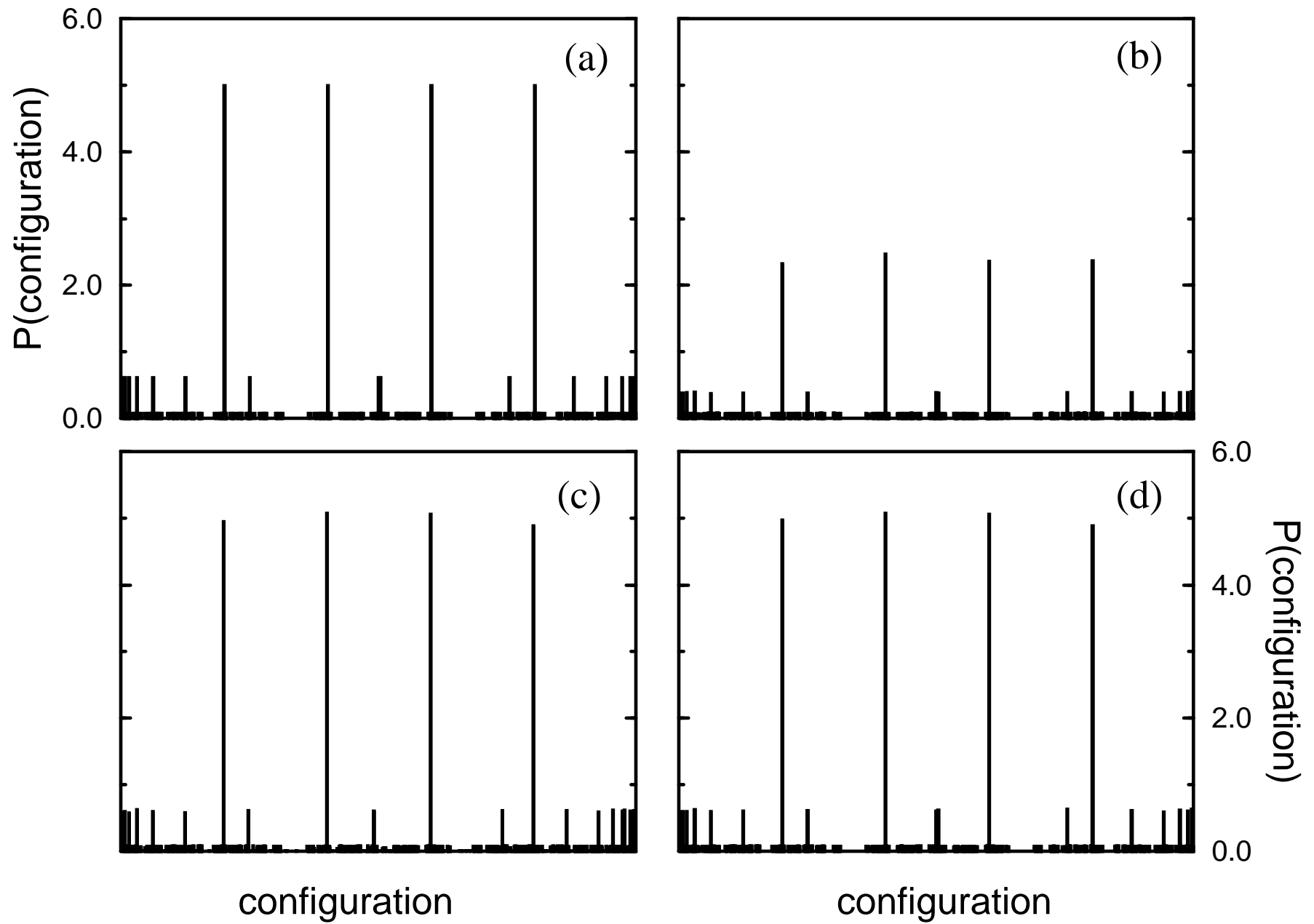
Fig 1. For $L=4$ there are 12870 spin configurations compatible with the
zero magnetization thermodynamic state. The figure shows the probability
distribution for the different configurations (arbitrarily numbered) at
$t=1.9$ ($t=kT/J$) according to (a) the Boltzmann distribution, (b) opposite
nearest neighbour pair exchange, (c) nearest neighbour pair exchange,
independently of sign and (d) opposite pair exchange at random distance. The
figure illustrates the fact that opposite spin exchange (case b) does not
lead to the correct distribution.

Fig 2. Two possible configurations of a fixed magnetization state for $L=5$.
If only opposite pairs are considered for exchange, transition $a\rightarrow
b$ will be selected for a trial move with probability $\frac 16$, while
transition $b\rightarrow a$ will be selected with probability $\frac 18$.

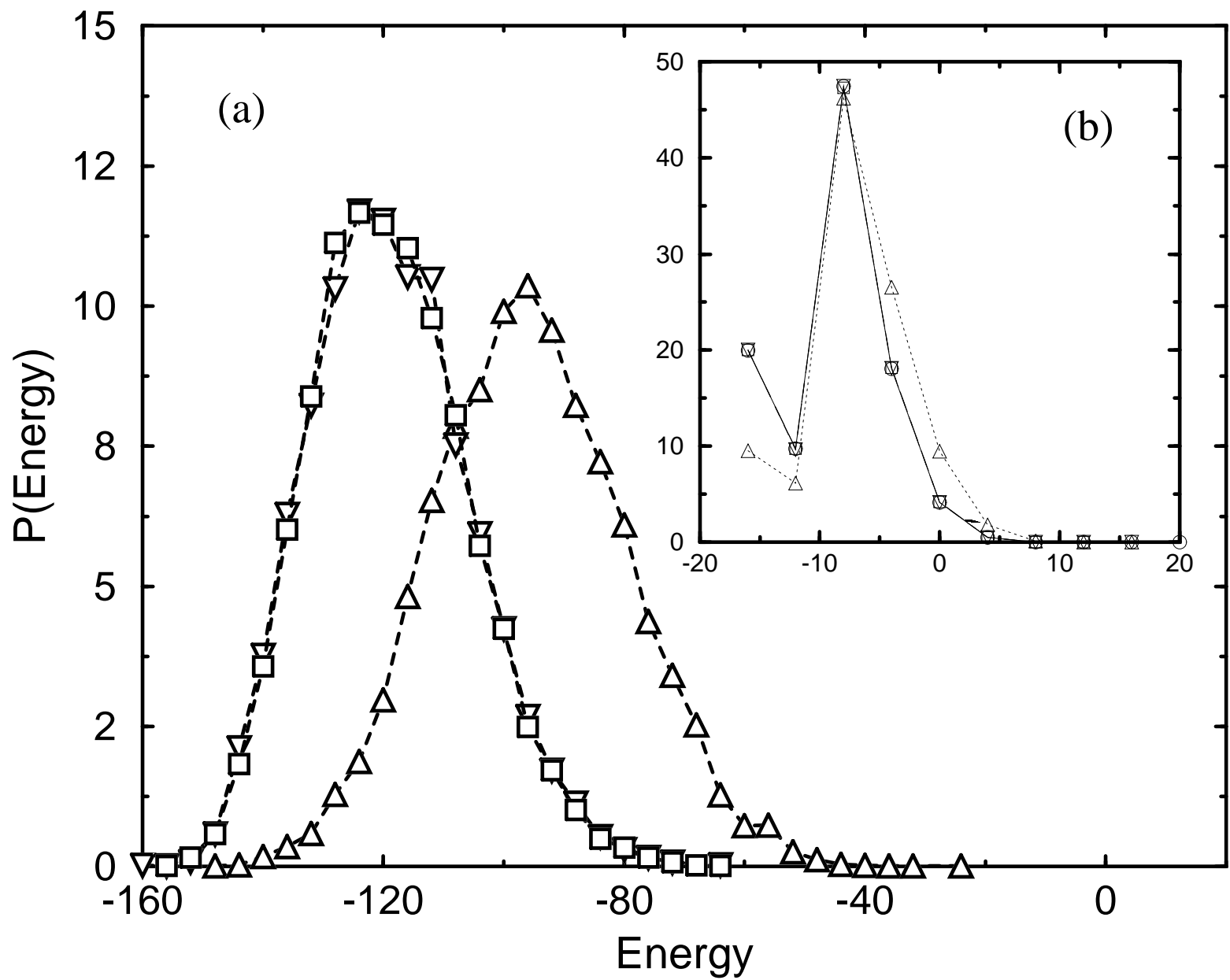
Fig 3. Energy probability distribution for (a) $L=10$ and (b) $L=4$ for the
reduced temperature $t=1.9$. ($\Delta $) nn opposite pair exchange, ($\nabla
$) nn pair exchange independent of sign and ($\Box $) random distance
opposite pair exchange results are presented in both figures. Boltzmann
distribution values ($\bigcirc $) are also presented for $L=4$ in fig (b).
Typically, 10$^6$ MC steps were used for the nearest neighbour exchange
simulations and 10$^5$ MC steps for the random distance algorithm.

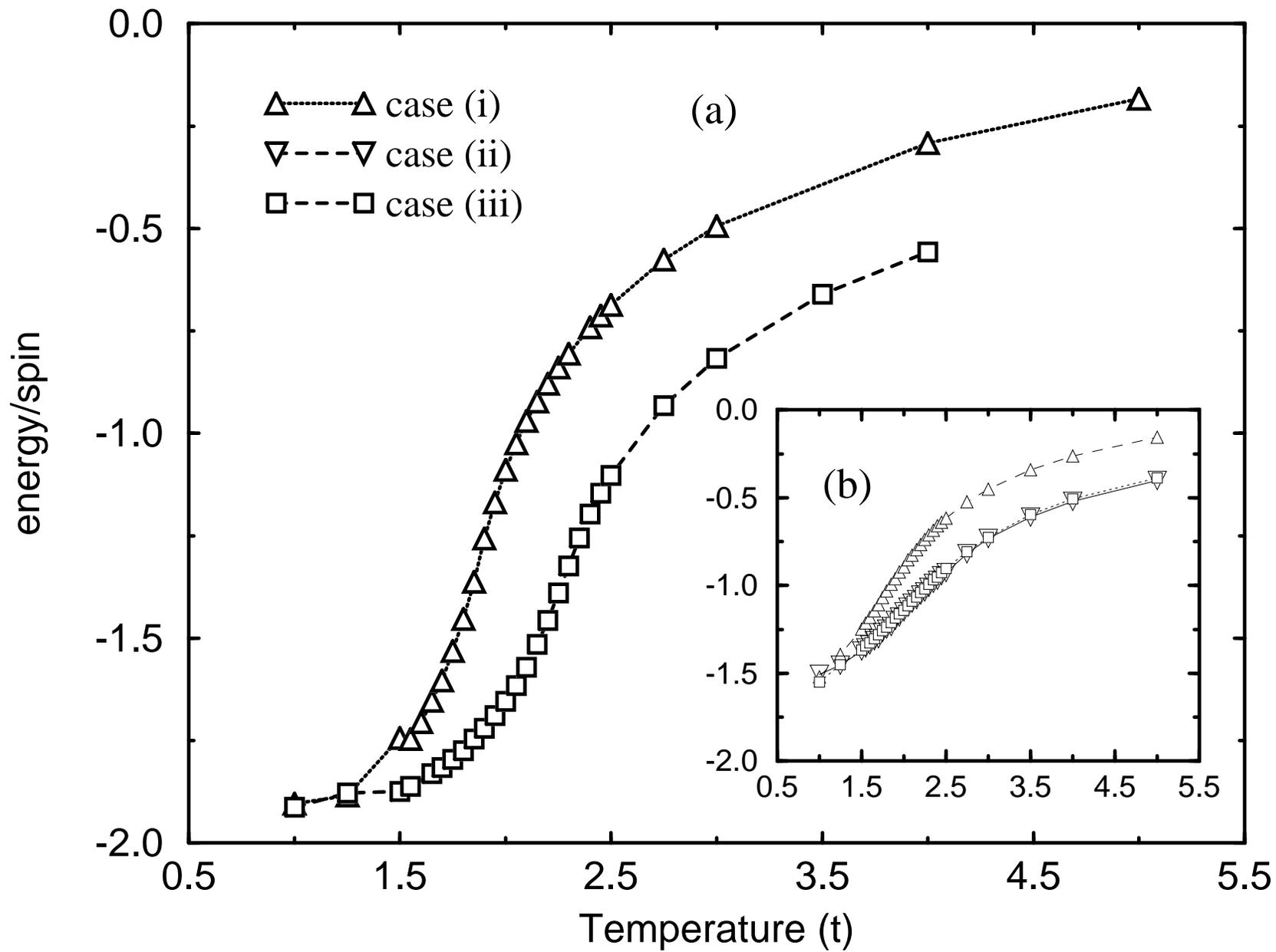
Fig 4. Energy per spin as a function of temperature for (a) $L=80$ and (b) 
$L=10$. Cases ($ii$) and ($iii$) (see text) coincide, whereas
case ($i$) would indicate a systematically larger average energy.

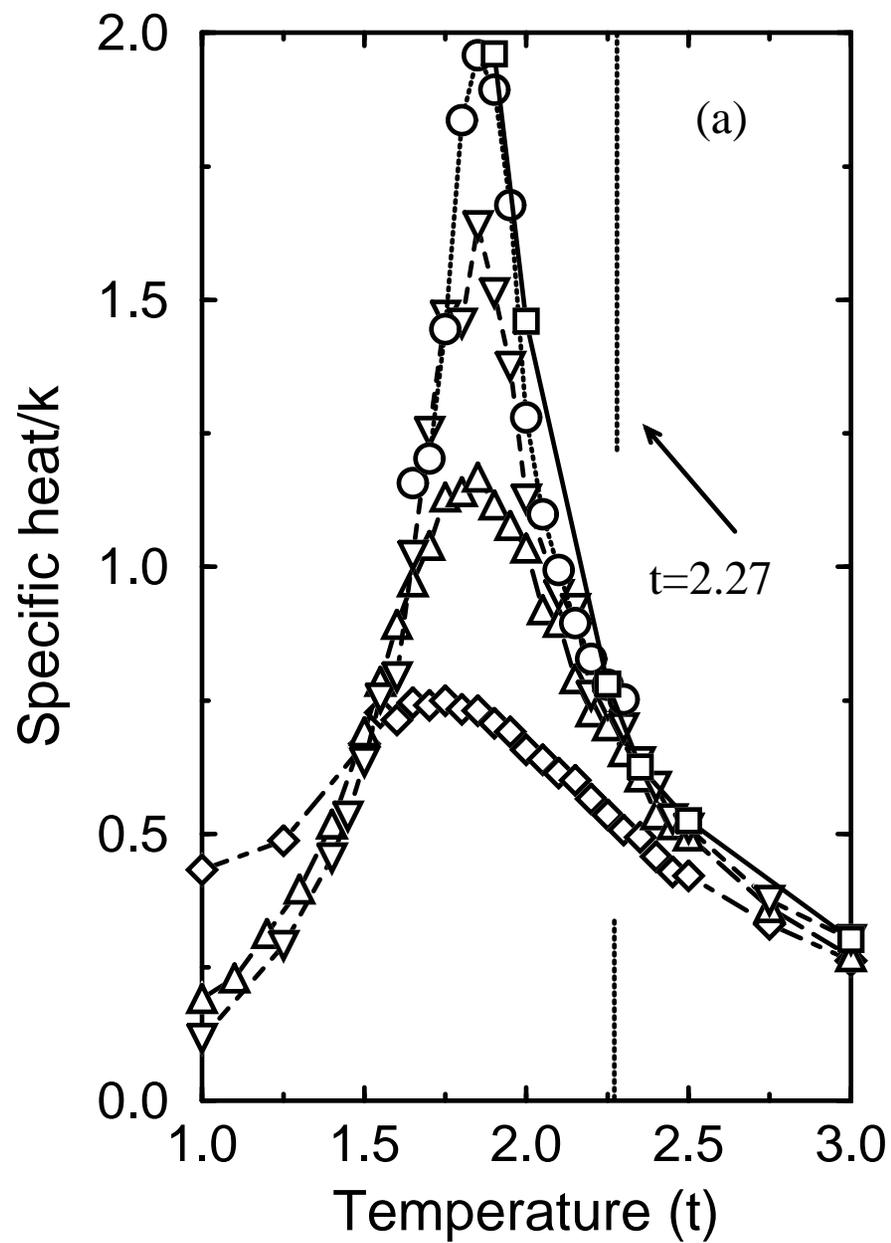
Fig 5. Specific heat as a function of temperature for different lattice
sizes (a) for opposite nn exchange and (b) for random distance opposite
exchange. Typical runs are as in Figs 3 and 4. The exact transition
temperature ($t=2.27$) is also indicated.

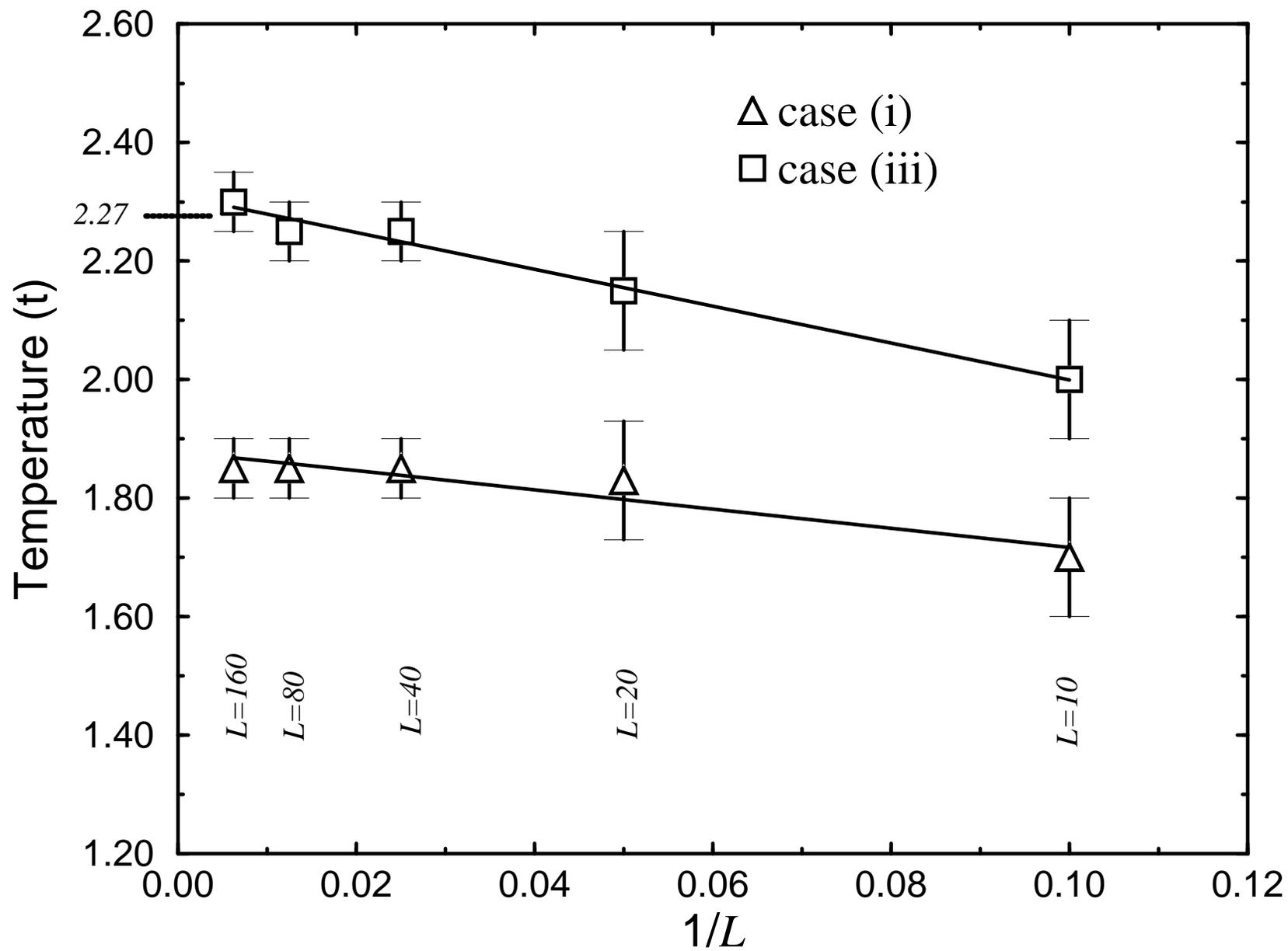
Fig 6. Energy as a function of time (in units of MC steps) for the $L=80$
lattice (a) from the nn opposite exchange algorithm for $t=1.95$ and (b)
from the random distance algorithm for $t=2.2$. These temperatures were
chosen in the region of large fluctuations, which differ for the two
algorithms. The figure shows that relaxation times (from an initial
disordered microstate) are of order 15$\times $10$^4$ for the first case and
5$\times $10$^3$ for the random distance case. Correlation times are
respectively of order 10$^3$ and 10 (not shown in the figure).


\begin{thebibliography}{99}
\bibitem{20}  In the general case, detailed balance will not be obeyed if
the number of opposite pairs changes after the transition. The algorithm
could be turned correct if one would multiply the Metropolis weight factor
by the ratio of the number of bonds for energy raising transitions.
\end{thebibliography}
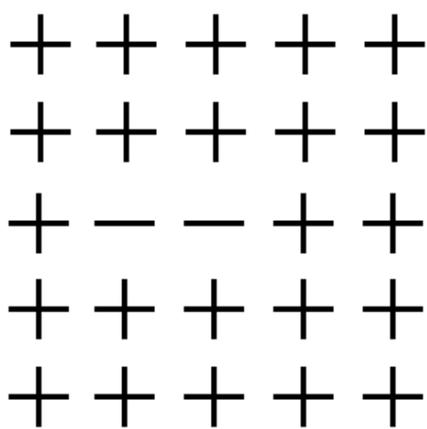
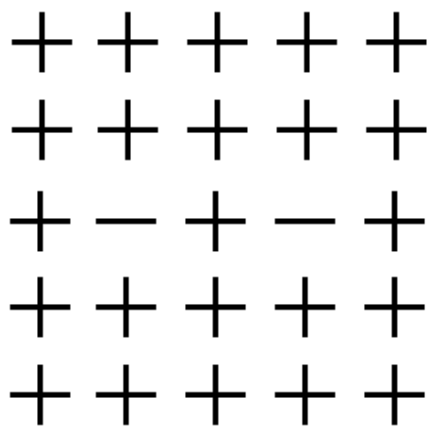
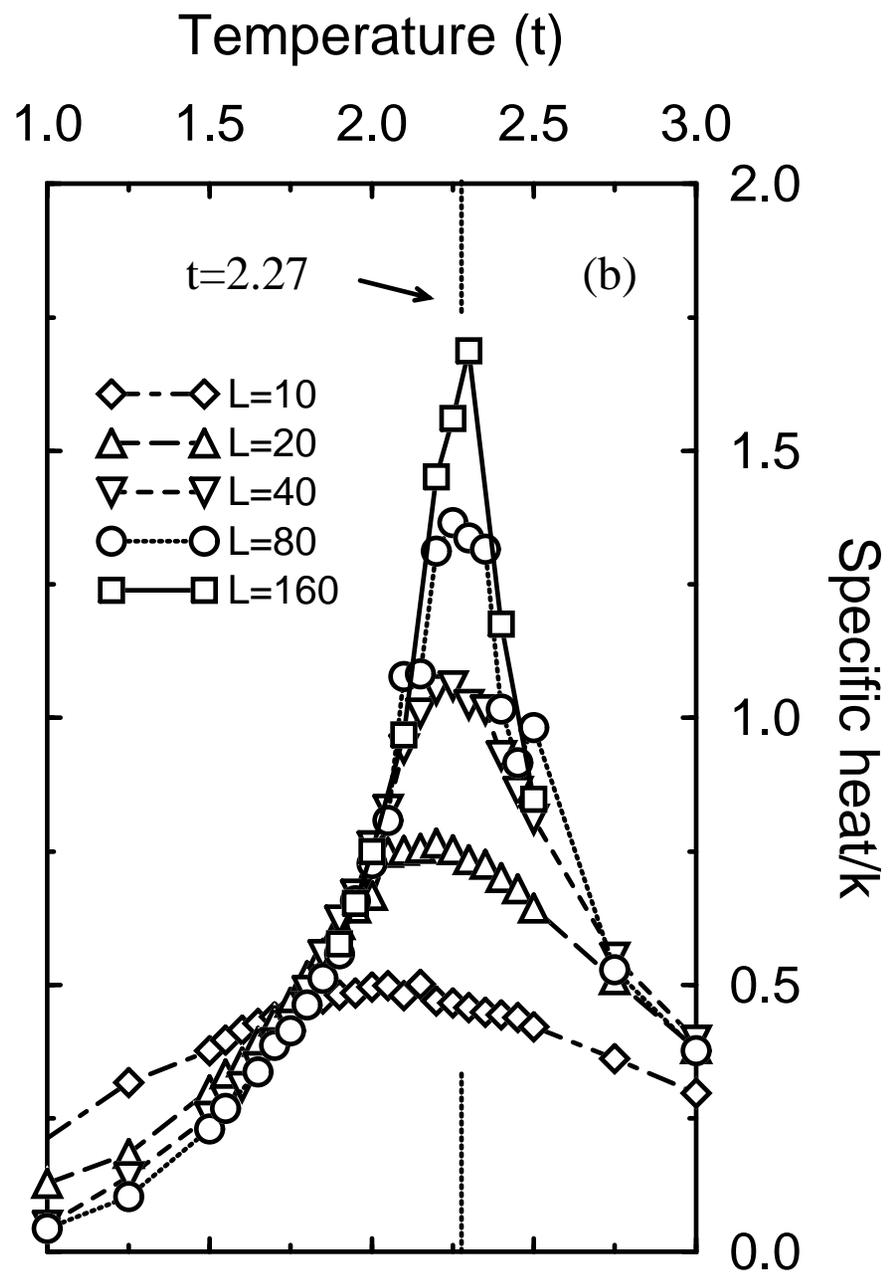
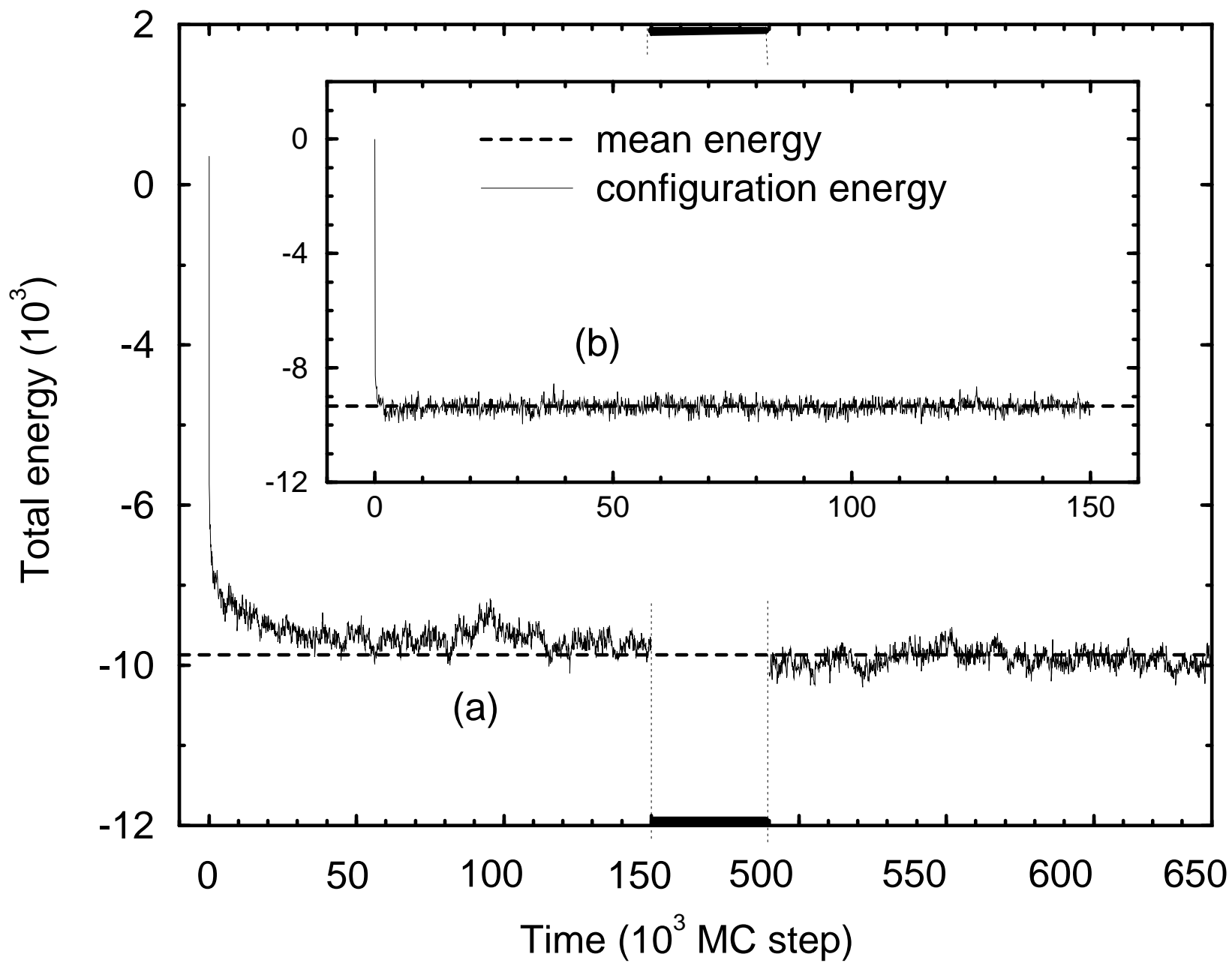
\end{document}